\documentclass{article}

\usepackage[preprint]{spconf}

\usepackage{amsmath,graphicx}
\usepackage{bm}
\usepackage{tikz}

\usepackage[ruled,vlined]{algorithm2e}
\usepackage{float}
\newcommand{\gvec}[1]{\bm{\mathit{#1}}}

\usepackage{tabularx}
\usepackage{booktabs}
\newcommand*\rot{\rotatebox{0}}
\usepackage{multirow}
\usepackage{multicol}

\newcommand\inputpgf[2]{{
		\let\pgfimageWithoutPath\pgfimage
		\renewcommand{\pgfimage}[2][]{\pgfimageWithoutPath[##1]{#1/##2}}
		\input{#1/#2}
}}

\title{Estimation of Microphone Clusters in Acoustic Sensor Networks using Unsupervised Federated Learning}
\name{Alexandru Nelus, Rene Glitza, and Rainer Martin\thanks{This work has been supported by the German Research Foundation (DFG) - Project Number 282835863.}}
\address{
    Institute of Communication Acoustics, Ruhr-Universit\"at Bochum, Bochum, Germany\\
    email: \texttt{\{firstname.lastname\}@rub.de}
}

\newcommand\copyrighttext{%
  \footnotesize \copyright 2021\ IEEE. Personal use of this material is permitted. Permission from IEEE must be obtained for all other uses, in any current or future media, including reprinting/republishing this material for advertising or promotional purposes, creating new collective works, for resale or redistribution to servers or lists, or reuse of any copyrighted component of this work in other works.}
   
\newcommand\copyrightnoticee{%
\begin{tikzpicture}[remember picture,overlay]
\node[anchor=south, yshift=15pt] at (current page.south) {\mbox{\parbox{\dimexpr\textwidth-\fboxsep-\fboxrule\relax}{\copyrighttext}}};
\end{tikzpicture}%
}

\begin{document}


\ninept
\maketitle

\copyrightnoticee

\begin{abstract}
In this paper we present a privacy-aware method for estimating source-dominated microphone clusters in the context of acoustic sensor networks (ASNs). The approach is based on clustered federated learning which we adapt to unsupervised scenarios by employing a light-weight autoencoder model. The model is further optimized for training on very scarce data. In order to best harness the benefits of clustered microphone nodes in ASN applications, a method for the computation of cluster membership values is introduced. We validate the performance of the proposed approach using clustering-based measures and a network-wide classification task.

\end{abstract}
\section{Introduction}
\label{sec:intro}

The continuously decreasing cost of acoustic sensors and the rise in popularity of wireless networks and mobile devices have helped to establish the technological infrastructure needed by (wireless) acoustic sensor networks (ASNs). These are useful in a larger number of applications, ranging from smart-homes and ambient-assisted living \cite{cobos2016acoustic} to machine fault diagnosis \cite{madhu2019robust} and surveillance \cite{park2019sound}.

Typically, ASN applications have to deal with multiple audio sources being active at the same time. In many such scenarios, the benefits provided by ASNs can be enhanced by having access to information regarding microphone nodes and their relation to the active acoustic sources. An example is provided in \cite{Gergen_journal_2015, Gergen_beamforming_2018} where the estimation of source-dominated microphone clusters based on hand-engineered spectral feature representations helps improve the ASN's overall signal classification and source separation performance.

While clustering and processing of multiple microphones will provide additional benefits to the aforementioned applications, the transmission of data-rich signal representations in a potentially unsafe (wireless) communication environment also poses serious privacy risks. Even in a small-scale scenario such as a smart-home, there are privacy hazards posed by eavesdroppers connecting to the network and intercepting data \cite{eavesdropping}. Moreover, in a world where privacy concerns have taken center-stage \cite{papernot2018sok} and privacy policies like the European Union General Data Protection Regulation (EU GDPR) \cite{GDPR} aim to enforce principles such as "privacy-by-design", a more privacy-sensible solution is recommended. Therefore, this work steers away from using feature representations derived directly from the raw audio data in favor of a more privacy-aware solution based on clustered federated learning (CFL) \cite{Sattler9174890}. In this way, ASN nodes (clients) need only share locally learned neural network parameter updates with a central node (server). Furthermore, by exploiting the cosine-similarity measure between parameter update vectors, a hierarchical clustering of clients can be achieved. 

Federated learning (FL) \cite{KonecnyMRR16, McMahanMRA16} and inherently CFL have been designed for massively distributed systems that handle large amounts of data and have been so far only used in (semi-) supervised learning applications where (weak) classification labels were available. However, in our ASN scenarios, e.g., a smart-home, clustering has to be performed on relatively short audio segments and, most importantly, without access to training labels. Thus, the adaptation of CFL to this unsupervised scenario and its implementation in the context of ASNs becomes a challenging task. To approach this topic, we like to extend and study CFL in the context of ASNs with two simultaneously active acoustic sources in a shoebox room. 

The remainder of this paper is structured as follows: we first discuss the relation to prior work, followed by a description of the proposed methods. We then continue by detailing the experimental scenarios and the results and finalize with conclusions and outlook.
\vspace{-0.2cm}

\section{RELATION TO PRIOR WORK}
\label{sec:prior}
\vspace{-0.2cm}

The estimation of source-dominated microphone clusters in ASN-related scenarios has been previously explored using various approaches and with different target applications. These have ranged from using coherence models \cite{himawan2010clustering} and energy decay information to eigenvectors \cite{bahari2017distributed} and spectral features \cite{madhu2019robust, Gergen_journal_2015, Gergen_beamforming_2018, DBLP:conf/iwaenc/GergenMM18}. In conjunction with unsupervised fuzzy clustering, the latter works obtain robust clustering and subsequently improved signal classification results. Despite the mentioned advantages, no privacy-preserving component is included, thus falling short of modern privacy requirements \cite{GDPR,DING2019129}. Moreover, supervised fuzzy clustering requires prior knowledge about the number of sources  \cite{Zadeh65}. 

Our proposed approach builds upon the concepts introduced by the aforementioned works and focuses on adding a privacy-preserving layer using a variation of FL \cite{KonecnyMRR16, McMahanMRA16}, namely CFL \cite{Sattler9174890}. Spectral features which, although aggregated still retain privacy-sensitive information \cite{NelusGM17}, are replaced with more privacy-aware, locally learned, neural network parameter updates. This approach reduces privacy risks considerably on its own and even more in conjunction with additional encryption,  differential privacy \cite{YangLCT19}, or encoding schemes \cite{SattlerMWS20}. To solve the unsupervised clustering tasks using CFL's privacy advantages, we propose using a light-weight autoencoder in each sensor node that permits unsupervised training. CFL will then compare and cluster the local updates of these distributed autoencoders in a central server and assign nodes to acoustic sources. To handle training with scarce data and to further reduce the communication overhead that arises from transmitting deep neural network (DNN) model updates \cite{SattlerWMS20, SattlerWMS19}, we limit the number of transmitted parameters and update only a part of the autoencoder.
\vspace{-0.1cm}


\begin{algorithm}[t!]
    \SetAlgoLined
    \KwIn{Pre-trained autoencoder $h$, thresholds $\epsilon_1$, $\epsilon_2$ and $\epsilon_3$, maximum no. of rounds $max_{\tau}$ }
     freeze all parameters of $h$ except $\gvec{\theta}$\\
     \While{\text{audio buffer $!=$ empty}}{
        read audio $\gvec{D}$ of $M$ clients\\
        initialize cluster list $C = \{\{1,..M\}\}$ with a single cluster element that contains all $M$ clients\\
        $C' = \{\}$\\
        $\gvec\theta_c \leftarrow$ \text{random initialization}\\
        \For{$\tau = 1 \ {\normalfont \textbf{to}} \ max_{\tau}$}{
             \For{$c \in C$}{
                 \For {$i \in c$}{
                 $\gvec{\Delta \theta}^{\tau}_{i} \leftarrow SGD(h_{\gvec{\theta}_{c}}(\gvec{D}_i))$\\
                 }
                 $\Delta\bar{\theta}_c = \left\| \frac{1}{|c|} \sum_{i \in c} \gvec{\Delta\theta}_i \right\| $\\
                 $\Delta\hat{\theta}_c = \max_{i \in c} (\| \gvec{\Delta\theta}_i \|)$\\
                      \eIf{$\Delta\bar{\theta}_c \leq \epsilon_1$ \text{\&} $\Delta\hat{\theta}_c \geq \epsilon_2$ \text{\&} $|\nabla \Delta\bar{\theta}_c| \leq \epsilon_3$}{
                        $a_{i,j} = \frac{\langle \gvec{\Delta\theta}_i, \gvec{\Delta\theta}_j \rangle}{\lVert \gvec{\Delta\theta}_i \rVert \lVert \gvec{\Delta\theta}_j \rVert}$, $\forall i,j \in c$\\
                        $c_1, c_2 \leftarrow $ bi-partition ($\gvec{A}$)\\
                        $\gvec{\theta}^{\tau+1}_{c_1} = \gvec{\theta}_{c}^{\tau} + \sum_{i\in c_1}\frac{|\gvec{D}_{i}|}{|\gvec{D}_{c_1}|}\gvec{\Delta\theta}^{\tau}_{i}$\\
                        $\gvec{\theta}^{\tau+1}_{c_2} = \gvec{\theta}_{c}^{\tau} + \sum_{j\in c_{2}}\frac{|\gvec{D}_{j}|}{|\gvec{D}_{c_2}|}\gvec{\Delta\theta}^{\tau}_{j}$\\
                        $C' = C' + \{c_1, c_2\}$\\
                        $\tau = max_{\tau} + 1$
                      }
                      {
                        $\gvec{\theta}_{c}^{\tau+1} = \gvec{\theta}_{c}^{\tau} + \sum_{i \in c}\frac{|\gvec{D}_{i}|}{|\gvec{D}_{c}|}\gvec{\Delta\theta}^{\tau}_{i}$\\
                        $C' = C' + \{c\}$\\
                      }
            }
            $C=C'$
        }
     }
     \caption{Unsupervised CFL for the estimation of source-dominated microphone clusters in ASNs}
     \label{alg:CFL}
\end{algorithm}

\vspace{-0.7cm}
\section{Unsupervised Clustering Using CFL}
\label{sec:methods}

\vspace{-0.1cm}
\subsection{Federated learning}
\label{sec:fl}

Federated learning was introduced in \cite{KonecnyMRR16, McMahanMRA16} as a method for large-scale privacy-preserving distributed learning of neural network parameters. It works using a three-step iterative process over a given number of communication rounds $\tau$. In the first step, the clients synchronize with the server by downloading the latest model parameters represented by column vector $\gvec{\theta}^{\tau}$. In the second step each client $i$ independently improves its own model parameters $\gvec{\theta}^{\tau}_{i}$ by performing stochastic gradient descent (SGD) \cite{bottou2010large} on their respective data $\gvec{D}_{i}$. In the third step, the clients upload their model parameters updates $\gvec{\Delta\theta}^{\tau}_{i}$ to the server for aggregation following
\vspace{-0.2cm}

\begin{equation}
    \gvec{\theta}^{\tau+1} = \gvec{\theta}^{\tau} + \sum_{i=1}^{M}\frac{|\gvec{D}_{i}|}{|\gvec{D}|}\gvec{\Delta\theta}^{\tau}_{i},
\end{equation}
 where $M$ is the total number of clients, $\gvec{D}$ their total dataset, and $|\cdot|$ denotes the cardinality of a dataset.

\vspace{-0.2cm}

\subsection{Clustered federated learning}
\label{sec:cfl}

It is shown in \cite{Sattler9174890, SattlerMWS20} that for the cases where clients' data comes from different (\textit{incongruent}) distributions, there is no single $\gvec{\theta^*}$ that can optimally minimize the loss of all clients at the same time. For this reason, the authors suggest clustering the clients that have similar (\textit{congruent}) distributions and training separate server models for each resulting cluster.
The clustering criterion proposed uses the cosine similarity measure $a_{i,j}$ between the nodes' weight update vectors following
\vspace{-0.1 cm}
\begin{equation}
    a_{i,j} = \frac{\langle \gvec{\Delta\theta}_i, \gvec{\Delta\theta}_j \rangle}{\lVert \gvec{\Delta\theta}_i \rVert \lVert \gvec{\Delta\theta}_j \rVert},
\end{equation}
where $\langle\cdot,\cdot\rangle$ denotes the inner product and $\lVert \cdot \rVert$ the $L_2$ norm. The cosine similarities $a_{i,j}$ of all clients are collected in the symmetric matrix $\gvec{A}$.

Hierarchical clustering using bi-partitioning can be recursively applied using $\gvec{A}$. The resulting two clusters $c_1$ and $c_2$ of each bi-partitioning step are derived such that the maximum cross-cluster cosine similarity is always smaller than the minimum of either intra-cluster cosine similarities \cite{Sattler9174890}:
\vspace{-0.08 cm}
\begin{equation}
    \max_{\forall i \in c_1, k \in c_2}(a_{i, k}) < \min(\min_{\forall i,j \in c_1}(a_{i, j}), \min_{\forall k,l \in c_2}(a_{k, l})).
    \vspace{-0.1 cm}
\end{equation}
The process is recursively repeated, and new sub-clusters are obtained, until the data distributions' congruence condition is no longer violated. The latter can be verified for each cluster $c$ by analyzing the mean and the maximum Euclidean norms of the weight update vectors $\gvec{\Delta\theta}_c$, defined as
\vspace{-0.2 cm}
\begin{equation}
    \Delta\bar{\theta}_c = \left\| \frac{1}{|c|} \sum_{i \in c} \gvec{\Delta\theta}_i \right\| \quad\mathrm{and}\quad \Delta\hat{\theta}_c = \max_{i \in c} (\| \gvec{\Delta\theta}_i \|).
    \vspace{-0.1 cm}
\end{equation}
Whenever the server has reached a stationary solution but some clients are still converging towards a locally stationary point, a low value of $\Delta\bar{\theta}_c$ in conjunction with a higher value of $\Delta\hat{\theta}_c$ is observed. This indicates incongruent data distributions and prompts bi-partitioning.

\subsection{Unsupervised clustered federated learning}

In other works, FL and CFL have been used for improving a server-based classification model with the goal of high classification accuracy. Our work, however, is only concerned with obtaining good clustering results which can be further used to enhance subsequent ASN-based applications. Moreover, we aim for a more general solution that does not rely on the availability of labeled data, thus requiring an unsupervised approach in the clustering process. As such, we propose to use a light-weight autoencoder with a low number of trainable parameters that are periodically re-initialized.

The proposed adaptation, along with the standard CFL algorithm \cite{Sattler9174890}, are schematically described in Algorithm 1. Prior to performing FL, the light-weight autoencoder $h$ is pre-trained, after which all layers except the bottleneck layer are frozen. The latter is always re-initialized with random parameters before applying CFL. The reduction of trainable network parameters is necessary in order to avoid overfitting \cite{ZhangBHRV17} caused by the large discrepancy between the very small number of training samples and the large number of parameters of a complete model. Moreover, processing fewer parameters reduces computational and bandwidth costs \cite{SattlerWMS20}.

Additional to the incongruity verification based on \mbox{$\Delta\bar{\theta}_c \leq \epsilon_1$} and $\Delta\hat{\theta}_c \geq \epsilon_2$ introduced in \cite{Sattler9174890}, we propose a supplementary verification as to fit the high $\Delta\bar{\theta}_c$ and $\Delta\hat{\theta}_c$ variation generated by handling a small number of training samples. This consists in \mbox{thresholding} the gradient $|\nabla \Delta\bar{\theta}_c| \leq \epsilon_3$, based on the intuition that a small $\Delta\bar{\theta}_c$ slope indicates the system reaching a stationary solution regardless of the absolute values of $\Delta\bar{\theta}_c$.

\subsection{Membership values}

\begin{figure*}[t]
    \centering

    \resizebox{0.95\textwidth}{!}{%
		\input{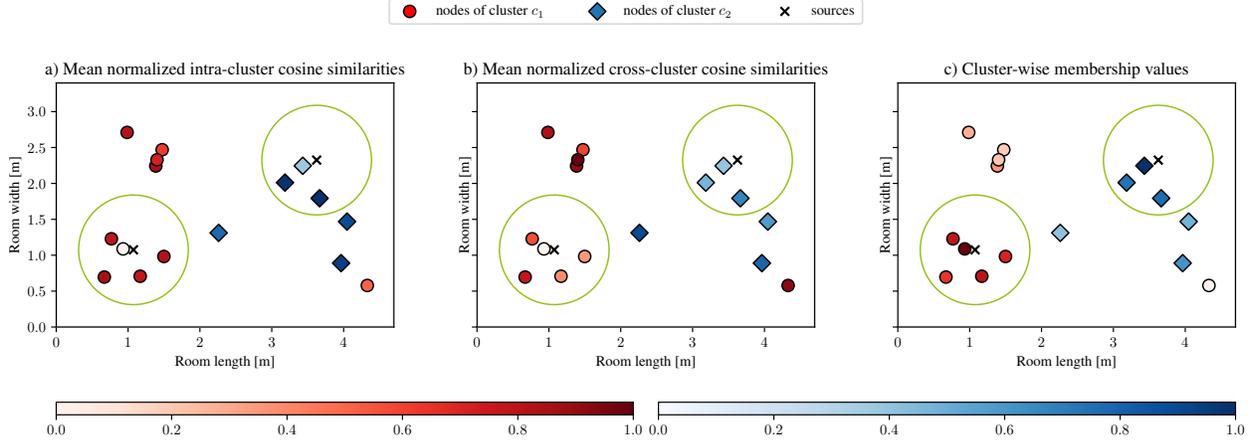}
	}
	\vspace{-0.3cm}
    \caption{Cluster membership values (MVs) for unsupervised CFL where mean normalized intra- (a) and cross-cluster cosine similarities (b) are aggregated in order to determine cluster reference nodes which in turn are used to compute MVs (c). Outliers are occasionally observed due to specific acoustic constellations which include wall reflections and reverberation. Illustration regards a single simulation scenario.}
    \label{fig:scenarios}
    \vspace{-0.29cm}
\end{figure*}

\vspace{-0.2cm}

To assess the contribution of each node to its respective cluster we propose the computation of cluster membership values (MVs) after each bi-partitioning of clients into clusters $c_1$ and $c_2$. We first compute the mean intra- and cross-cluster similarities for each client $i$ and stack them in vectors $\gvec{q}$ and $\gvec{r}$, respectively, with
\begin{equation}
    q_i = \frac{1}{|c_x|-1} \sum_{j \in c_x \setminus - \{i\}} a_{i,j} \enskip\mathrm{and}\enskip r_i = \frac{1}{|c_y|}  \sum_{k \in c_y} a_{i,k}
\end{equation}
for $\forall i \in c_x$ and $(c_x, c_y) \in \{(c_1, c_2), (c_2, c_1)\}$, where $|\cdot|$ denotes the cardinality of a set.
We further apply min-max normalization to $\gvec{q}$ and $\gvec{r}$ after which we compute vector $\gvec{p}$ that contains the aggregated mean cosine similarity values for each client using a weighted sum,
\begin{equation}
\label{eq:mv}
    p_i = \lambda \frac{q_i - \min(\gvec{q})}{\max(\gvec{q})-\min(\gvec{q})} + (1-\lambda) \frac{r_i - \min(\gvec{r})}{\max(\gvec{r})-\min(\gvec{r})}.
\end{equation}
 Since the acoustic sources are modeled as spherical point sources, nodes very close to a source pick up quite different signals from other nodes which are dominated by a reverberant mixture of both sources. This results in small mean intra-cluster similarity values for nodes positioned close to the cluster source and those positioned at extremities, thus requiring additional cross-cluster information in order to distinguish them. After applying (\ref{eq:mv}), only the nodes closest to a cluster source will display small $p_i$ values. We further select the node with the smallest $p_i$ value in each cluster as a reference node and compute the MVs vector $\gvec{\mu}$ as the cosine similarities between the cluster nodes and their respective reference node:
\begin{equation}
    \label{eq:argmin}
    \mu_{i} = a_{i,\arg \min (p_j)}, \forall i,j \in c_x \enskip\mathrm{and}\enskip  c_x \in \{c_1, c_2\} \, .
\end{equation}
Min-max normalization is again applied to vector $\gvec{\mu}$. An example of a single simulation scenario is provided in Figure \ref{fig:scenarios}. Additionally, thresholding with $\mu_i=0, \forall \mu_i \leq v$ is also considered in order to disregard nodes with low MVs.
\vspace{-0.2cm}

\section{Experiments and results}
\label{sec:experiments}

\subsection{Database and simulation scenarios}
\label{sec:dataset}

For this work, we employ a subset of the LibriSpeech corpus \cite{panayotov2015librispeech}, namely \textit{train-clean-100}, which consists of 251 speakers (125 female, 126 male) from audiobook recordings sampled at 16 kHz. We further apply voice activity detection (VAD) and restructure the data into 25006 utterances of length 10 s each. Next, the dataset is split into \textit{Libri-server} with 157 speakers (79 female, 78 male)   used to train the autoencoder and gender recognizer and \textit{Libri-clients} containing 94 speakers used to perform clustering and infer speaker genders.

The current work only considers two simultaneously active sources randomly positioned in opposing quadrants of a shoebox room of size $4.7 \times 3.4 \times 2.4$ m with the reverberation time $T_{60}=0.34$ s. The ASN deployed in the room consists of $M=16$ microphone nodes, which are, as well, randomly spread under the constraint that for every source, a minimum of three nodes is positioned within critical distance, thus having higher direct component energy than reverberation energy. The random spread of sources and microphones is performed ten times. For each constellation created, 20 gender-balanced speaker pairs are randomly selected from Libri-clients resulting in 200 simulation scenarios. For each scenario, we randomly select 16 utterances/speaker to perform CFL followed by gender recognition using the estimated cluster configuration.

Each ASN node $i$ is exposed to a mix of signals from both sources $s_1$ and $s_2$, expressed as 
\vspace{-0.1cm}
\begin{equation}
\label{eq:IRs}
    x_i(t) = s_1(t)*g_{i}^{s_1}(t) + s_2(t)*g_{i}^{s_2}(t),
\vspace{-0.1cm}
\end{equation}
where $g_{i}^{s_j}$ is the impulse response from source $j$ to node $i$ and is simulated with CATT Acoustic using cone-tracing \cite{DalenbaeckTUCT2019}.

\vspace{-0.2cm}
\subsection{Server pre-training}
\label{sec:pretraining}

The architecture of the proposed autoencoder $h$ is detailed in Table \ref{tab:autoencoder}. This is trained to reconstruct the Log-Mel Band Energy (LMBE) input feature representation $\gvec{Y}$. The latter is extracted for each 10 s utterance as detailed in \cite{Nelus2019variational} using a short-time discrete Fourier transform (STFT) with window length $L_1=0.064$ s and step size $R_1=0.032$ s along with $K=128$ Mel filters. Training is performed on the Libri-server set for 300 epochs using an SGD optimizer with a learning rate of $l_r=0.1$. The loss function that is minimized over the entire model parameters set $\gvec{\Theta}$  is the mean squared error (MSE) between the input and reconstructed feature vectors:
\vspace{-0.22cm}
\begin{equation}
\label{eq:MSE}
    \min_{\gvec{\Theta}}L_{\text{mse}}(\gvec{Y, \hat{Y}}) = \min_{\gvec{\Theta}}\frac{1}{N} \sum_{n=1}^{N}(y_{n}-{\hat y_{n}})^2.
\vspace{-0.2cm}
\end{equation}
After the model is trained, the $\gvec{\Theta}$ parameters, except for subset $\gvec{\theta}$, are frozen. The initial number of $O_1 = 5999$ trainable parameters is thus reduced to $O_2 = 841$. This subset corresponds to the parameters of Layer 5 from Table \ref{tab:autoencoder}. These are further re-initialized and trained as described in the next subsection.
\vspace{-0.3cm}


\subsection{Clustering}
\label{sec:clustering}

\begin{table}[t]
\centering
\caption{Neural network architecture of autoencoder $h$.}
\vspace{0.15cm}
\label{tab:autoencoder}
\resizebox{\columnwidth}{!}{%
\begin{tabular}{c|c|c|c|c|c|c}
\hline
Layer & Input         & Operator    & \begin{tabular}[c]{@{}c@{}}Out\\ ch.\end{tabular} & Stride & \begin{tabular}[c]{@{}c@{}}Kernel/\\ Nodes\end{tabular} & Activation \\ \hline
1     & 128 x 128     & Conv2d      & 6                                                 & 1      & 5 x 5                                                   & ReLu       \\
2     & 6 x 124 x124  & MaxPool     & -                                                 & 2      & 2 x 2                                                   & -          \\
3     & 6 x 62 x 62   & Conv2d      & 16                                                & 1      & 5 x 5                                                   & ReLu       \\
4     & 16 x 58 x 58  & MaxPool     & -                                                 & 2      & 2 x 2                                                   & -          \\ \hline
5     & 16 x 29 x 29  & Dense       & -                                                 & -      & 29                                                      & ReLu       \\ \hline
6     & 16 x 29 x 29  & Unpool      & -                                                 & 2      & 2 x 2                                                   & -          \\
7     & 16 x 58 x 58  & ConvTrans2d & 6                                                 & 1      & 5 x 5                                                   & ReLu       \\
8     & 6 x 62 x 62   & Unpool      & -                                                 & 2      & 2 x 2                                                   & -          \\
9     & 6 x 124 x 124 & ConvTrans2d & 1                                                 & 1      & 5 x 5                                                   & Sigmoid   
\end{tabular}%
}
\end{table}

\begin{table}[t]
\vspace{-0.4cm}
\centering
    \caption{Normalized cluster-to-source distance $\Tilde{d}_{c_x}^{s_z}$ from cluster $c_x$  to source $s_z$, averaged over 200 scenarios.}
    \vspace{0.15cm}
    \label{table:cluster2source}
    \begin{tabular}{l|ll}
    \cline{1-3}
        & \multicolumn{1}{l|}{$c_1$} & $c_2$   \\ \hline
    $s_1$ & \textbf{0.15}                    & 0.91 \\ \hline
    $s_2$ & 0.87                    & \textbf{0.14}
    \end{tabular}%
    \vspace{-0.1 cm}
\end{table}

The pre-trained autoencoder $h$ detailed in Subsection \ref{sec:pretraining} is employed for unsupervised CFL as indicated in Algorithm \ref{alg:CFL}. The MSE loss function introduced in (\ref{eq:MSE}) acts only on parameters subset $\gvec{\theta}$, as the rest of the parameters are frozen. We empirically set $\epsilon_1=0.0134$, $\epsilon_2=0.005$, $\epsilon_3=0.0007$, and $max_\tau=25$ communication rounds. Each of the total $M=16$ clients trains for one epoch in every round $\tau$, with $l_r=0.1$. The clients' input feature representation $\gvec{Y}_i$ is computed based on the 10 s time-domain signal $x_i(t)$ as detailed in (\ref{eq:IRs}), where $s_1$ and $s_2$ are selected from the Libri-clients dataset.
The MVs defined in (\ref{eq:mv}) are computed using $\lambda=0.5$.

To evaluate and compare the clustering performance to state-of-the-art solutions \cite{Gergen_journal_2015}, we employ from the latter the normalized cluster-to-source distance from cluster $c_x$ to source $s_z$ as 
\begin{equation}
    \Tilde{d}_{c_x}^{s_z} = \frac{\lVert \rho_{s_z} - \bar{\rho}_{c_x} \rVert}{\lVert \rho_{s_1} - {\rho}_{s_2} \rVert}, \forall c_x \in \{c_1, c_2\} \enskip\mathrm{and}\enskip s_z \in \{s_1, s_2\}, 
\end{equation}
 where $\rho_{s_z}$ is the geometric position of source $s_z$ and $\bar{\rho}_{c_x}$ is the average of geometric positions of nodes $i$ assigned to cluster $c_x$ which are weighted by their respective MVs. Table \ref{table:cluster2source} shows $\Tilde{d}_{c_x}^{s_z}$ averaged over all 200 simulation scenarios.
 
It can be observed that low values of $\Tilde{d}_{c_1}^{s_1}$ and $\Tilde{d}_{c_2}^{s_2}$ are obtained, indicating that the MV-weighted cluster centers are each situated in close proximity to their dominant source. The high $\Tilde{d}_{c_1}^{s_2}$ and $\Tilde{d}_{c_2}^{s_1}$ values validate that the MV-weighted cluster centers are, concurrently, situated away from their opposing sources. The obtained metrics point towards good cluster predictions, comparable to results in \cite{Gergen_journal_2015}. 
\vspace{-0.2cm}

\subsection{Gender recognition}
\label{sec:genderrecog}


\begin{table}[t]
    \footnotesize
	\centering
	\caption{Neural network architecture of gender recognizer $e$.}
	\vspace{0.15cm}
	\resizebox{\columnwidth}{!}{%
		\begin{tabular}{@{}cccccccc@{}}
			\toprule
			\rot{Input}                      & \rot{Operator}                      & \rot{\begin{tabular}[c]{@{}c@{}}Out \\ ch.\end{tabular}} & \rot{Stride}                 & \rot{\begin{tabular}[c]{@{}c@{}}Kernel/ \\ Nodes\end{tabular}} & \rot{\begin{tabular}[c]{@{}c@{}}Batch \\ norm.\end{tabular}} & \rot{Activation}                   & \rot{Dropout} \\ \midrule

			\multicolumn{1}{c|}{501$\times$40} & \multicolumn{1}{c|}{Conv2d}  & \multicolumn{1}{c|}{32}                                       & \multicolumn{1}{c|}{1} & \multicolumn{1}{c|}{5$\times$40}                                        & \multicolumn{1}{c|}{Yes}                                       & \multicolumn{1}{c|}{ReLU}    & -    \\

			 \multicolumn{1}{c|}{497$\times$32} & \multicolumn{1}{c|}{MaxPool} & \multicolumn{1}{c|}{-}                                        & \multicolumn{1}{c|}{1} & \multicolumn{1}{c|}{5$\times$1}                                         & \multicolumn{1}{c|}{-}                                         & \multicolumn{1}{c|}{-}       & -       \\

			\multicolumn{1}{c|}{99$\times$32}  & \multicolumn{1}{c|}{Conv2d}  & \multicolumn{1}{c|}{64}                                       & \multicolumn{1}{c|}{1} & \multicolumn{1}{c|}{3$\times$32}                                        & \multicolumn{1}{c|}{Yes}                                       & \multicolumn{1}{c|}{ReLU}    & -    \\

			 \multicolumn{1}{c|}{92$\times$64}  & \multicolumn{1}{c|}{MaxPool} & \multicolumn{1}{c|}{-}                                        & \multicolumn{1}{c|}{1} & \multicolumn{1}{c|}{92$\times$1}                                        & \multicolumn{1}{c|}{-}                                         & \multicolumn{1}{c|}{-}       & -       \\

			\multicolumn{1}{c|}{1$\times$64}   & \multicolumn{1}{c|}{Dense}    & \multicolumn{1}{c|}{-}                                        & \multicolumn{1}{c|}{-} & \multicolumn{1}{c|}{64}                                         & \multicolumn{1}{c|}{-}                                         & \multicolumn{1}{c|}{ReLU}    & 50\%       \\


			 \multicolumn{1}{c|}{1$\times$64}    & \multicolumn{1}{c|}{Dense}    & \multicolumn{1}{c|}{-}                                        & \multicolumn{1}{c|}{-} & \multicolumn{1}{c|}{2}                                           & \multicolumn{1}{c|}{-}                                         & \multicolumn{1}{c|}{Softmax} & - \\      
		\end{tabular}%
	}
	\label{table:gender}
\end{table}

In order to assess the clustering performance from a utility standpoint, we further implement a gender recognition task. To this end, we first train a gender recognition model $e$ on the Libri-server dataset, where a part of the clean signals is randomly augmented with male-female reverberant signal mixtures. The model architecture is described in Table \ref{table:gender}, where the input feature representation $\gvec{Y}$ is extracted for $L_1=0.064$, $R_1=0.02$, and $K=40$. Training is performed for 13 epochs using a cross-entropy loss function and an SGD optimizer with $l_r=0.01$. Testing is performed for all 200 simulation scenarios indicated above, utilizing the utterances from Libri-clients used to generate clustering estimations.

The proposed evaluation metrics are Accuracy ($A_{cc}$) and $F_1$-score ($F_1$), where the ground truth gender label of a node is given by the gender of the source with the shortest first peak delay of the impulse response. The ground truth gender label of a cluster is given by the mode of the ground truth gender labels of its constituting nodes. 
The predicted gender label of a cluster is the mode of predicted gender labels of its nodes. Since each node processes 16 utterances/scenario, its gender label is given by the mode of gender predictions across the utterances. The evaluation metrics are averaged across all 200 simulation scenarios. Results are shown in the first column of Table \ref{table:accuracy}.

The aforementioned results are further improved by taking into account the nodes' dominant-source proximity indicated by the MVs. In this case, the predicted gender label of a cluster is given by the sum of MV-weighted node predictions normalized by the sum of MVs. In the second column of Table \ref{table:accuracy}, it is observed how the MV-weighting has a positive effect on gender recognition performance as smaller/larger distances between sources and nodes imply less/more reverberation and signal interference, thus leading to more/less accurate node-wise gender predictions.

Moreover, in an additional experiment, the MVs smaller than a threshold $v$ are set to 0 to exclude poorly performing nodes. The threshold is systematically varied and results are shown in the second to last columns of Table \ref{table:accuracy}. The results from the previous experiment where no thresholding was used correspond to $v=0$. It is observed that for an increasing $v$, gender recognition scores gradually improve. This, in conjunction with the previous results, indicates that the proposed clustering approach has a significant performance-enhancing effect on a network-wide task.

\begin{table}[t]
\vspace{-0.5cm}
\centering
\caption{Aggregated gender recognition Accuracy ($A_{cc}$) and $F_1$-score ($F_1$) of estimated clusters, without and with membership value (MV) weighting using threshold $v$. Results reflect 200 scenarios.}
\vspace{0.15cm}
\label{table:accuracy}
\begin{tabular}{c|clcc}
       & \begin{tabular}[c]{@{}c@{}}no \\ MV\end{tabular} & \begin{tabular}[c]{@{}l@{}}MV\\ $v$=0\end{tabular} & \begin{tabular}[c]{@{}c@{}}MV\\ $v$=0.5\end{tabular} & \begin{tabular}[c]{@{}c@{}}MV\\ $v$=0.9\end{tabular} \\ \hline
        $A_{cc}$(\%)  & 90 & 96  & 97 & 99 \\ \hline
        $F_1$(\%) & 89 & 96 & 97 & 98                                                
\end{tabular}%
\vspace{-0.17cm}
\end{table}

\vspace{-0.1cm}
\section{Conclusions and outlook}
\label{sec:conclusions}
\vspace{-0.1cm}

We have proposed an unsupervised adaptation of CFL to ASN scenarios by using a light-weight autoencoder as server and isolating a subset of its parameters for re-initialization and re-training in FL rounds. An additional bi-partitioning indicator was introduced along with a novel method for generating cluster membership values. It has been empirically demonstrated that the presented privacy-aware approach offers good clustering performance by means of cluster-to-source distance measures and the performance of a multi-sensor gender recognition task. A more comprehensive investigation using a larger variety of acoustic conditions along with a more detailed assessment of privacy risks is planned for future works.





\bibliographystyle{IEEEbib}
\bibliography{refs}

\end{document}